\documentclass[acus]{JAC2001}

\usepackage{graphicx}
\usepackage{epsfig}

\begin{document}

\thispagestyle{empty}
\onecolumn
\begin{flushright}
{\small
SLAC--PUB--8884\\
June 2001\\}
\end{flushright}

\vspace{.8cm}

\begin{center}
{\large\bf
Frequency Resolved Measurement of Longitudinal Impedances Using
Transient Beam Diagnostics\footnote{Work supported by
Department of Energy contract  DE--AC03--76SF00515.}}

\vspace{1cm}

D. Teytelman, J. Fox, S. Prabhakar\\
Stanford Linear Accelerator Center, Stanford University,
Stanford, CA  94309\\

\medskip

J. Byrd\\
Lawrence Berkeley National Laboratory,
1 Cyclotron Road, Berkeley, CA  94563\\
 
\end{center}

\vfill

\begin{center}
{\large\bf
Abstract }
\end{center}

\begin{quote}
In this paper we present several techniques for characterizing
longitudinal impedances based on transient measurements of the growth rates
and tune shifts of unstable coupled-bunch modes. These techniques are
applicable to measurement of both fundamental and higher-order mode
impedances and allow characterization of shunt impedances and quality
factors of the HOMs. Methods presented here are complementary to lab bench
measurements of RF cavities, in that the beam based measurements directly
sense the physical impedance in the installed configuration. In contrast to
a single-bunch integrated impedance measurement these techniques resolve the
impedances in the frequency domain. These methods allow determination of the
impedance's unaliased frequency by analyzing synchronous phase
transients. Experimental results from ALS and BESSY-II are presented showing
the use of these techniques to measure complex impedances.
\end{quote}

\vfill

\begin{center}
{\it Presented at IEEE Particle Accelerator Conference (PAC 2001), Chicago,
Illinois, 18-22 Jun 2001} \\
\end{center}

\newpage
\pagenumbering{arabic}
\pagestyle{plain}
\twocolumn

\title{Frequency Resolved Measurement of Longitudinal Impedances Using
Transient Beam Diagnostics\thanks{This work was supported by DOE contract
No. DE-AC03-76SF00515}
}

\author{D. Teytelman\thanks{dim@slac.stanford.edu}, J. Fox, S. Prabhakar
Stanford Linear Accelerator Center, Stanford, CA\\
J.Byrd, Lawrence Berkeley National Laboratory, One Cyclotron Road, Berkeley,
CA}
\maketitle

\begin{abstract}
In this paper we present several techniques for characterizing
longitudinal impedances based on transient measurements of the growth rates
and tune shifts of unstable coupled-bunch modes. These techniques are
applicable to measurement of both fundamental and higher-order mode
impedances and allow characterization of shunt impedances and quality
factors of the HOMs. Methods presented here are complementary to lab bench
measurements of RF cavities, in that the beam based measurements directly
sense the physical impedance in the installed configuration. In contrast to
a single-bunch integrated impedance measurement these techniques resolve the
impedances in the frequency domain. These methods allow determination of the
impedance's unaliased frequency by analyzing synchronous phase
transients. Experimental results from ALS and BESSY-II are presented showing
the use of these techniques to measure complex impedances.
\end{abstract}

\section {Introduction}

The interaction of charged particles in a storage ring or circular
accelerator with the ring impedance determines many important accelerator
dynamics parameters. Single and multi-bunch instabilities are the result of
interactions of the bunches with the impedance of the machine, and achieving
high stored currents requires knowledge and control of the ring components
which produce the dominant narrow-band impedances. There are several
laboratory techniques to measure impedances of physical components
\cite{Palumbo:1989tr,Corlett:1993iw}. Beam-based impedance measurement
techniques exist as well. Frequency-resolved information about the coupling
impedance can be extracted from a measurement of the beam transfer function
(BTF) \cite{Hofmann:1977km}. However such a measurement can only be
performed below the instability threshold. In addition network analyzer
sweeps have to be repeated for each unstable mode making BTF approach slow
and cumbersome.

This paper presents several beam-based longitudinal impedance measurement
techniques. These fast transient multi-bunch techniques measure the aliased
longitudinal impedance as a function of frequency in a sampling bandwidth up
to 1/2 the RF frequency. Consequently various higher-order mode resonators
can be identified and their complex impedance (and parameters such as center
frequency and Q) measured.

\section {Longitudinal impedances and coupled-bunch instabilities}
\label{sec:instabilities}

Bunches of charged particles passing through the vacuum chamber of a storage
ring leave behind electromagnetic fields. These fields (wake fields) affect
the energy of the following bunches providing a bunch-to-bunch coupling
mechanism. At high beam currents such coupling can cause instabilities.

The bunch motion in a storage ring can be projected onto the orthonormal
basis of the even fill eigenmodes (EFEMs).
Eigenvalue of mode $l$ is given by \cite{Prabhakar:2000wp}

\begin{eqnarray}
\Lambda_l = -d_r + j\omega_s + 
\frac{\pi \alpha e f_{rf}^2 I_0}{E_0 h \omega_s} 
Z^{\parallel eff}(l\omega_0 + \omega_s)
\label{eq:Z2gr}\\
Z^{\parallel eff}(\omega) = \frac{1}{\omega_{rf}} 
             \sum_{p=-\infty}^{\infty} (ph\omega_0+\omega)
             Z^{\parallel}(ph\omega_0 + \omega)
\label{eq:Zeff}
\end{eqnarray}
where $d_r$ is the radiation damping rate, $\omega_s$ is the synchrotron
frequency, $\alpha$ is the momentum compaction factor, $e$ is the charge of
the electron, $f_{rf}$ is the frequency in the accelerating cavities, $I_0$
is the beam current, $E_0$ is the beam energy, $h$ is the ring harmonic
number, $\omega_0$ is the revolution frequency, and $Z^{\parallel}(\omega)$
is the total longitudinal impedance.

In order to measure modal eigenvalues $\Lambda$ we use the capabilities of a
programmable longitudinal feedback system \cite{Fox:1999ig}. The system is
able to measure the unique synchronous phase and centroid motion of every
bunch in a storage ring, and uses digital memory to record time sequences of
the bunch motion. In a transient grow/damp measurement feedback loop is
opened under software control for a predetermined period of time and then
closed. In the open-loop conditions unstable modes grow exponentially due to
noise and feedback system records the motion of the bunches during the
transient. The motion is then projected on the EFEM basis and modal
exponential growth and damping rates as well as oscillation frequencies are
extracted \cite{Prabhakar:1999ij}. Once the eigenvalues are measured it is
possible to extract the aliased impedance according to Eq.~\ref{eq:Z2gr}.
The aliased beam-derived impedance, combined with knowledge about the
impedances from bench measurements of ring components may be properly
assigned as an unaliased impedance.

\section{Synchronous phase transients}
\label{sec:synchro}

For the cases when ring fill pattern is uneven additional information about
the impedance can be obtained from analyzing the dependence of
synchronous phases on bunch currents. Previous work by Prabhakar
\cite{Prabhakar:1998jp} presents the relationship between the bunch
currents, impedances, and synchronous phases. This work is applicable to
fill patterns where all buckets are populated, however unevenly. For empty
buckets synchronous phase is not measurable. Extending the analysis to fills 
with empty buckets (gaps) we get

\begin{eqnarray}
\vec{\phi}_U = \frac{-N}{|V_c cos(\phi_s^0)|} {\bf A}^{UV}
\vec{Z}^{\dagger}_V
\label{eq:reduced}\\
Z^{\dagger}_n = \sum_{m=-\infty}^{\infty}Z^{\parallel}((m N + n)\omega_0)
\nonumber
\end{eqnarray}

where $\vec{\phi}$ is the vector of bunch phases, $U$ is the set that
includes all non-empty buckets, $V_c$ is the peak RF cavity voltage and
$\phi_s^0$ is the synchronous phase in absence of wake fields. Matrix 
${\bf A}^{UV}$ is computed using inverse DFT (Discrete Fourier Transform)
matrix and a DFT of the vector of individual bunch currents. Set $V$
includes revolution harmonics excited by the DFT of bunch currents. By
solving an overdetermined linear system of equations described by
Eq.~\ref{eq:reduced} in the least-squares sense we obtain
$\vec{Z}^{\dagger}_V$.

\section{ALS measurements}

The goal of the first measurement is to quantify the HOM impedances of the
two 500~MHz main RF cavities installed at the ALS. Past measurements have
determined that there are two dominant EFEMs, modes 205 and 233, excited by
the impedances in the main RF cavities \cite{Prabhakar:1997xa}. Using the
measurements made on the spare cavity identical to the ones installed in the
ring mode 205 had been identified as driven by the $TM_{011}$ longitudinal
mode at 812~MHz. Mode 233 has two potential driving HOMs, at 2.353~GHz and
2.853~GHz \cite{Corlett:1993iw}.

Due to technical limitations it is only possible to fill 320 RF
buckets at the ALS. All of the transient measurements described here were
taken with 320 buckets maximally equally filled leaving a gap of 8 RF
buckets. Since the gap is small we assume that eigenmodes of the fill are
very close to those of an even-fill.

\begin{figure}[htb]
\centering
\includegraphics*[width=73mm]{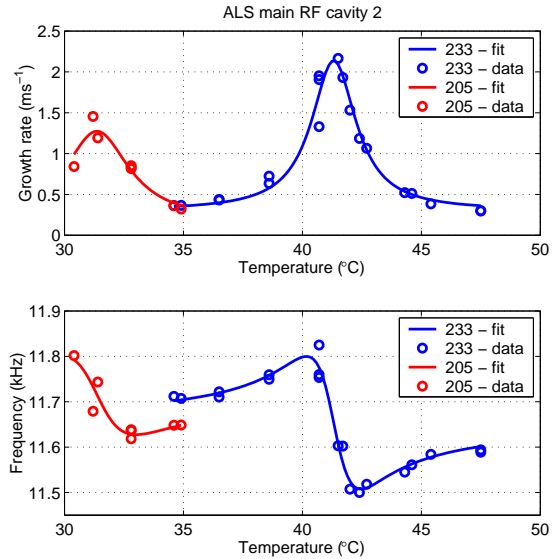}
\caption{Growth rates (top) and oscillation frequencies (bottom) of modes
205 and 233 in main RF cavity 2 normalized to 100~mA.}
\label{fig:cav_T}
\end{figure}

In order to characterize the frequency dependence of the impedance we
shifted the center frequencies of the cavity HOM resonances by changing the
temperature of the cavity. At each
point the temperature was allowed to stabilize and the open-loop eigenvalues
of the unstable modes were measured using the transient grow-damp technique.
In Fig.~\ref{fig:cav_T} the growth rates and oscillation frequencies of
modes 205 and 233 are plotted versus temperature of cavity 2.

\begin{table}[htb]
\begin{center}
\caption{
\label{table:ALS}
Extracted HOM parameters for ALS RF cavities}
\vspace{1ex}
\begin{tabular}{lccc}
Cavity&1&2&2\\
\hline
$F_r$, GHz&2.8532&2.8532&0.8119\\
$R_s$, k$\Omega$&$55\pm2$&$97\pm3$&$210\pm20$\\
$Q,\ \times 10^3$&$21\pm2$&$24\pm2$&$12\pm3$\\
$R/Q$, $\Omega$&$2.6\pm0.2$&$4.0\pm0.3$&$17\pm4$
\end{tabular}
\end{center}
\end{table}

These measurements agree well with the expected effect of the HOM resonators.
However these measurements do not provide a way to distinguish between the
two possible HOMs at 2.353 and 2.853~GHz as the source of the aliased
impedance. To resolve this ambiguity the ring was filled with
a single bunch while a cavity probe signal was monitored on a spectrum
analyzer. We observed that change of cavity temperature had very small
effect on the magnitude of the revolution harmonics excited within the
2.353~GHz resonance while signal at 2.853~GHz scaled with temperature in
agreement with the growth rate measurements.
Thus the resonance measured in the temperature scan is at 2.853~GHz.
In addition the impedance presented by the 2.353~GHz HOM can be considered 
constant.

In order to quantify impedance parameters $R_s$ and $Q$ we convert cavity
temperatures to center frequencies of the resonance. Conversion factor is
determined by matching cavity probe signal levels between two temperatures and two
RF frequency settings. Using nonlinear least-squares estimation we extract
parameter values. In Table~\ref{table:ALS} results for both cavities are
summarized. Note that characteristics of the 2.853~GHz resonances in two
cavities differ significantly. The cavities have RF windows of different
designs which can cause variations in the $R/Q$ values. Additionally, the
mode in question is close to the beam pipe cut-off frequency and is strongly
affected by the field leakage.

Using growth rates vs. RF cavity temperature results it is possible to
optimize operating temperatures of the main RF cavities. Since temperatures
affect the transverse impedances as well as longitudinal impedances, mapping
growth rates in horizontal and vertical planes is necessary for a full
understanding of the tradeoff.

\section{BESSY-II measurements}

These measurements were aimed at quantifying longitudinal impedances at
BESSY-II. The machine was filled with 350 consecutive
bunches out of 400 to a current of 165~mA. A series of 15 transient
grow/damp experiments was conducted over a period of 10 minutes during which
the machine configuration remained unchanged. There are three unstable EFEMs
seen in the data: 281, 396, and 397. Using Eq.~\ref{eq:Z2gr} we extract
complex longitudinal impedances from the measured growth rates and
oscillation frequencies.

\begin{eqnarray*}
Z^{\parallel eff}_{281} &=& (63.2\pm8.1) + (0\pm94)j\ \textrm{k}\Omega\\
Z^{\parallel eff}_{396} &=& (59.0\pm3.3) + (1115\pm53)j\ \textrm{k}\Omega\\
Z^{\parallel eff}_{397} &=& (59.6\pm3.7) - (726\pm36)j\ \textrm{k}\Omega
\end{eqnarray*}

Impedance measurement for modes 396 and 397 correlates well with the
impedance of four third harmonic cavities parked between 3 and 4 revolution
harmonics below $3f_{rf}$.

\begin{figure}[htb]
\centering
\includegraphics*[width=83mm]{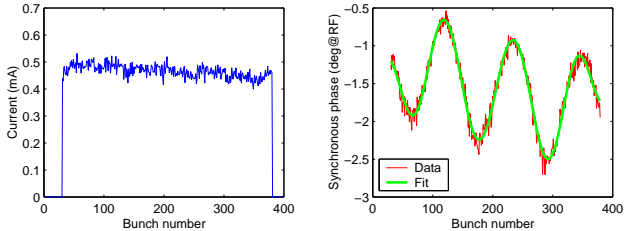}
\caption{Bunch-by-bunch currents (left) and synchronous phases (right)
extracted from BESSY-II dataset.}
\label{fig:bessy_synchp}
\end{figure}

As described in Sec.~\ref{sec:synchro} we can estimate the impedance by
analyzing the synchronous phase transient. In Fig.~\ref{fig:bessy_synchp}
synchronous phase transient in BESSY-II is presented with
350 consecutive buckets filled nearly equally.
Periodic pulse excitation of the fill pattern generates oscillatory behavior
of the synchronous phases. Solving
Eq.~\ref{eq:reduced} in the least-squares sense we obtain aliased
impedances. Least-squares estimate of the synchronous phases is also shown
in Fig.~\ref{fig:bessy_synchp} for comparison with experimental data. Using
15 BESSY transient measurements described above we get 
$Z^{\dagger}_{396} = (35 \pm 22) + (344 \pm 14)j$ k$\Omega$ and 
$Z^{\dagger}_{397} = (22 \pm 6) - (233 \pm 15)j$ k$\Omega$.

The two methods of measuring the impedance can be used together in order to
determine unaliased frequencies. This is possible due to the fact that
during aliasing into $Z^{\parallel eff}$ impedance is scaled by resonant
frequency, while in $Z^{\dagger}$ it is unscaled. From Eq.~\ref{eq:Zeff} we
have

\begin{eqnarray}
\label{eq:aliasing}
|Z^{\parallel eff}_{396}| &=& \frac{(p h + 396) \omega_0}{h \omega_0}
|Z^{\dagger}_{396}|\\
\nonumber
p_{exp} &=& \frac{|Z^{\parallel eff}_{396}|}{|Z^{\dagger}_{396}|} - 
\frac{396 \omega_0}{400 \omega_0} = 2.2
\end{eqnarray}

Since $p$ in Eq.~\ref{eq:aliasing} is an integer by definition, comparison
above indicates that the physical impedance is at $2 \omega_{rf} + 396
\omega_0 = 3 \omega_{rf} - 4 \omega_0$. This conclusion agrees perfectly
with the expected position of the parked third-harmonic cavities.

\section{Summary}
\label{sec:summary}

We have demonstrated several methods for measuring the impedance of
accelerator components using transient diagnostic capabilities of the
DSP-based longitudinal feedback systems. The methods extend the capabilities
of laboratory bench measurements by quantifying the physical impedances as
installed in the accelerator. Dependence of the impedances on operating
conditions such as temperature or tuner position can be extracted and used
to select optimal working points. By comparing information obtained from
growth transients with the analysis of the synchronous phase transients for
uneven fills it is possible to determine the spectral position of the
driving impedance.

\section{Acknowledgments}

Authors would like to thank Jorn Jacob of ESRF and Greg Stover of LBNL for
help in setting up and conducting ALS measurements. We also thank Shaukat
Khan and Tom Knuth for setting up and taking BESSY-II transient measurements.

\end{document}